\documentclass[prd,aps,nofootinbib,showpacs,showkeys]{revtex4}
\usepackage{graphicx,epsf,amsfonts,amssymb,amsbsy}
\newcommand{\ds}{\displaystyle}
\newcommand{\vev}[1]{\langle#1\rangle}
\newcommand{\vect}{\left ( \begin{array}{c}}
\newcommand{\evect}{\end{array} \right )}

\begin{document}

\title{Abnormal number of Nambu-Goldstone bosons in the color-asymmetric 2SC phase of
an NJL-type model}
\author{D. ~Blaschke}
\email{david.blaschke@physik.uni-rostock.de}
\affiliation{Fachbereich Physik, Universit\"at Rostock, D-18051
Rostock, Germany\\
Joint Institute for Nuclear Research, 141980 Dubna, Moscow Region,
Russia}
\author{D.~Ebert}
\email{debert@physik.hu-berlin.de}
\affiliation{Institut f\"ur Physik,
Humboldt-Universit\"at zu Berlin, D-12489 Berlin, Germany}
\author{K.\,G.~Klimenko}
\email{kklim@mx.ihep.su}
\affiliation{Institute
of High Energy Physics, 142281 Protvino, Moscow Region, Russia}
\author{M. K. Volkov, and V. L. Yudichev}
\email{yudichev@thsun1.jinr.ru}
\affiliation{Joint Institute for Nuclear Research, 141980 Dubna,
Moscow Region, Russia}

\begin{abstract}
We consider an extended Nambu--Jona-Lasinio model including both $(q
\bar q)$- and $(qq)$-interactions with two light-quark flavors
in the presence of a single (quark density) chemical potential.
In the color superconducting phase of the quark matter the color
$SU_c(3)$ symmetry is spontaneously broken down to $SU_c(2)$.
If the usual counting of Goldstone bosons would apply,
five Nambu--Goldstone (NG) bosons corresponding to the
five broken color generators should appear in the mass spectrum.
Unlike that expectation, we find only three gapless diquark
excitations of quark matter. One of them is an
$SU_c(2)$-singlet, the remaining two form an $SU_c(2)$-(anti)doublet
and have a quadratic dispersion law in the small momentum limit.
These results are in agreement with the Nielsen--Chadha theorem, according to which
NG-bosons in Lorentz-noninvariant systems, having a quadratic dispersion law,
must be counted differently. The origin of the abnormal number of
NG-bosons is shown to be related to a nonvanishing expectation value of the color charge
operator $Q_8$ reflecting the lack of color neutrality of the ground state.
Finally, by requiring color neutrality, two massive diquarks are argued to
become massless, resulting in a normal number of five NG-bosons with usual linear dispersion laws.

\end{abstract}

\maketitle


\section{Introduction}
It is well known that, in accordance with the Goldstone theorem
\cite{gold,kibble}, $N$ Nambu--Goldstone (NG) bosons appear in
Lorentz-invariant systems if an internal continuous symmetry
group ${\cal G}$ is spontaneously broken down to a subgroup
${\cal H}$ (here $N$ is the number of generators in the coset
space ${\cal G}/{\cal H}$), i.~e. the number of NG-modes is
equal to the number of broken generators. However, in
Lorentz-noninvariant systems the number of NG-bosons can be less
than $N$. In this case, the counting of NG-bosons is regulated by
the Nielsen--Chadha (NC) theorem \cite{nielsen}: Let $n_1$ and $n_2$
be the numbers of gapless excitations that in the
limit of long wavelengths  have the dispersion laws $E\sim |\vec p|$
and  $E\sim |\vec p|^2$, respectively, then $N\le n_1+2n_2$.
(Here, $E$ is the energy and $\vec p$ is the three-momentum of
the particle.) In particular, this theorem is valid for
relativistically covariant theories as well, since in this case:
i) the total number of NG-bosons equals  $N$, the number of
broken symmetry generators \cite{kibble}; ii) evidently, the
dispersion law for these $N$ massless excitations looks like
$E\sim |\vec p|$, thus $N=n_1$.

Recently, in some relativistic models describing the dynamics of
the kaon condensate in the color-flavor-locked phase of dense
quark matter, an abnormal number of NG-bosons has been revealed
\cite{ms,sss}. Since the  Lorentz invariance is broken in this
case and some of the gapless excitations have a quadratic dispersion
law, there are no contradictions with neither Goldstone nor NC
theorems. The superfluid $^3$He in the A-phase \cite{volovik}
and ferromagnets \cite{nambu,hof} are other known examples of
condensed-matter systems with an abnormal number of NG-bosons.

In the present paper, we demonstrate the abnormal number of
NG-bosons in the dense color superconducting phase (2SC) of
quark matter for a simple version of the Nambu--Jona-Lasinio (NJL) model with two
light quarks and a single (quark number) chemical potential.
In this phase, which can be realized naturally
only at sufficiently large values of the chemical potential (300 MeV$\leq \mu<$1 GeV) , the
initial color $SU_c(3)$  symmetry is spontaneously broken down
to the $SU_c(2)$ group.
Hence, in accordance with the usual counting of the
Goldstone theorem, one might expect five NG-bosons, corresponding to the
five broken symmetry generators, to
appear in the NJL model.  We shall prove that in the
2SC phase only three gapless excitations that can be identified with NG-bosons appear. Two of them have a
quadratic dispersion law, thus yielding no contradiction with the
NC theorem. For further applications, notice also the following
important criterion which is sufficient for the equality between the number of NG-bosons
and the number of broken generators \cite{sss}: If $Q_i, i=1,..., N$ is the full set of
broken generators and if $\vev{[Q_i,Q_j]}=0$ for any pair $(i,j)$, then the number of
NG-bosons is equal to the number $N$ of broken generators.

Recall, that there is an alternative approach to study
color superconductivity, which is based on the weak coupling perturbative QCD \cite{qcd}.
In this case, using the Schwinger-Dyson equation with one gluon exchange,
the color superconducting phase was proved to exist at asymptotically high densities.
In the two-flavored QCD investigations of color superconductivity
five scalar NG-bosons are shown to exist, which are mixing with gluons
and required as longitudinal components for five massive gluons
(color Meissner effect) \cite{msw}.
Hence, the results concerning the number of NG-bosons of both the QCD and the NJL model
approaches  seem to be in contradiction\footnote{Of course,
it is necessary to point out that the QCD weak coupling considerations
are valid at asymptotic values of the chemical potential,
i.~e.{} at $\mu>1$ GeV, whereas NJL model results are correct for
sufficiently lower values  $\mu<1$ GeV. Thus, a direct comparison of results seems
to be problematic.}.
The origin of the above discrepancy is related to the fact that the ground
state of the considered simple NJL--type model is not color-neutral.
Indeed, since the expectation
value of the color charge operator $Q_8$ is non-vanishing, $\vev{Q_8}\ne 0$,
the criterion \cite{sss} is not applicable, and an
abnormal number of NG-bosons must arise.
On the other hand, in quark models with a color-neutral
ground state, according to the above criterion, there should
arise a  normal number of five NG-bosons all having normal (linear) dispersion laws.
Most interestingly, we find from the analysis of the mass spectrum
in our model that the masses of two scalar diquarks
are proportional to  the ground state expectation value of the color charge $\vev{Q_8}$.
Thus, when color neutrality, i.~e.{} $\vev{Q_8}=0$,
is imposed in NJL--type models as an additional condition (which can be realized
by introducing a color-chemical potential, $\mu_8$, "by hand" \cite{huang}),
then two additional massless particles
should appear, and the total number of NG-bosons would be equal to five.
Recently, it was shown that the ground state
of the 2SC phase of QCD is automatically color-neutral \cite{rebhan} due to a
dynamical generation of the color chemical potential $\mu_8$ by gluon condensation.
So, we see that
the original discrepancy in the number of NG-bosons between the QCD approach and
an extended NJL approach would disappear.

The paper is organized as follows. In Section II we investigate the considered NJL--type model
in the framework of the Nambu--Gor'kov formalism and derive an effective meson--diquark action
for a finite chemical potential. Moreover, the gap equations for the quark and diquark condensates
are numerically studied.
Sections III and IV contain a detailed analysis of the massless and massive
excitations in the diquark sectors
and explicitly show the influence of the color properties of the ground state
on the diquark mass spectrum. Section V contains a summary  and discussions.
Finally, dispersion laws for gapless diquarks are derived in  Appendix.

\section{The model and its effective action}

It is well known that perturbative methods are not applicable in
the low-energy/density QCD regions. Instead,  effective field
theories are usually considered. The most popular effective theories are
based on  Lagrangians with four-fermion interactions, like the
NJL-type models. Let us first give several (very approximate)
arguments somehow justifying the chosen structure of our QCD-motivated NJL
model introduced below. For this aim, consider two-flavor QCD with
a nonzero chemical potential and the color group $SU_c(3)$.
By integrating in the generating functional of QCD over gluons
and further ``approximating'' the nonperturbative gluon propagator by
a $\delta-$function, one arrives at an effective local chiral
four-quark interaction of the NJL type describing low-energy hadron
physics. Finally, by performing a Fierz transformation of the
interaction term and taking into account only scalar and
pseudo-scalar $(\bar q q)$- as well as scalar $(qq)$-interaction channels,
one obtains a four-fermionic
model given by the following Lagrangian (in Minkowski space-time notation)%
\footnote{The most general Fierz transformed four-fermion
interaction includes additional vector and axial-vector $(\bar q
q)$ as well as pseudo-scalar, vector and axial-vector-like
$(qq)$-interactions. However, these terms are omitted
here for simplicity.}
\begin{eqnarray}
  L=\bar q[\gamma^\nu i\partial_\nu+  \mu\gamma^0-m_o]q+G_1[(\bar
  qq)^2+(\bar qi\gamma^5\vec \tau q)^2]+
G_2[\bar q^C\epsilon^\delta\gamma^5\tau^2 q]
  [\bar q\epsilon^\delta\gamma^5\tau^2 q^C].
  \label{1}
\end{eqnarray}
In (\ref{1}), $\mu\geq 0$ is the quark chemical potential,
which in isotopically symmetric quark matter is the same for both quark flavors,
$q^C=C\bar q^t$, $\bar q^C=q^t C$ are charge-conjugated spinors, and
$C=i\gamma^2\gamma^0$ is the charge conjugation matrix (the symbol
$t$ denotes the transposition operation). The quark field
$q\equiv q_{i\alpha}$ is a flavor doublet and color triplet as
well as a four-component Dirac spinor, where $i=1,2$; $\alpha =
1,2,3$. (Latin and Greek indices refer to flavor and color
indices, respectively; spinor indices are omitted.) Furthermore,
we use the notations  $\vec \tau\equiv (\tau^{1},
\tau^{2},\tau^3)$ for Pauli matrices in flavor space;
$(\epsilon^\delta)^{\alpha\beta}\equiv\epsilon^{\alpha\beta\delta}$
is the totally antisymmetric tensor in  color space, respectively.
Clearly,  Lagrangian (\ref{1}) is invariant under the chiral
$SU(2)_L\times SU(2)_R$  (at $m_o=0$) and color $SU_c(3)$
symmetry groups. The physics of light mesons
\cite{ebvol,volk,hatsuda,volkyud}, diquarks \cite{ebka,vog} and
meson-baryon interactions \cite{ebjur,reinh} was successfully
described in the framework of different NJL models. These
effective theories were involved for the investigation of both ordinary
\cite{asakawa,ebert,klim}, hot and dilute \cite{hatsuda}, and
color superconducting dense quark matter
\cite{alford,buballa,yudichev,bvy03}.
Usually, on the basis of light-meson and baryon phenomenology, the
following restrictions on the coupling constants are considered
\begin{equation}
G_1>\pi^2/(6\Lambda^2),~~~~ G_2<G_1~,
\label{1a}
\end{equation}
where $\Lambda$ is the cutoff parameter in the
three-dimensional momentum space, necessary to eliminate the
ultraviolet divergences appearing when quantum effects (loops)
are taken into account (usually $\Lambda<1$ GeV).
Since (\ref{1}) is a low-energy effective theory for QCD, the
external parameter $\mu$ is restricted from the above: $\mu<\Lambda$.
In the region of coupling constants
(\ref{1a}) and at small values of $\mu$ (the case of low baryon
densities), only the operator $\bar q q$ has a nonzero vacuum
expectation value. Thus,  the chiral
symmetry of the model is spontaneously broken in this case,
while $SU_c(3)$ remains intact. The behavior of quark and meson
masses is established rather well in quark matter of low density
\cite{asakawa,ebert}. In particular, the number of $\pi$-mesons,
which are the three NG-bosons in this phase, is equal to the
number of broken symmetry generators. Hence, this example is a
good demonstration of the fact that Lorentz noninvariance is a
necessary but not sufficient condition for the abnormal number
of NG-bosons to appear in the system.  At sufficiently high
values of the chemical potential $\mu\sim 300 \div 350$ MeV,
the two-flavor color superconductivity phase occurs in model
(1) \cite{yudichev}. So, a nonzero diquark condensate ($\vev{qq}\ne
0$) is formed in this case, and, as a consequence, color
symmetry is spontaneously broken down to the $SU_c(2)$ subgroup.
A naive counting gives us five NG-bosons in this case (it is
just the number of broken generators of the $SU_c(3)$ group).
However, as it will be shown further, there are only three
gapless bosonic excitations of the 2SC quark-matter ground
state. Two of them satisfy the quadratic dispersion law, in
agreement with the NC theorem.

The linearized version of Lagrangian (\ref{1}) that contains
auxiliary bosonic fields has the following form
\begin{eqnarray}
\tilde L\ds &=&\bar q[\gamma^\nu i\partial_\nu +\mu\gamma^0
 -\sigma-m_o -i\gamma^5\vec
 \tau\vec\pi]q-\frac{\sigma^2+\vec \pi^2}{4G_1}-
 \nonumber\\ &-&\frac{\Delta^{*\delta}\Delta^\delta}{4G_2}+
\frac{i\Delta^{*\delta}}{2}[\bar q^C\epsilon^\delta\gamma^5\tau^2 q]
 -\frac{i\Delta^\delta}{2}[\bar q \epsilon^\delta\gamma^5\tau^2 q^C].
 \label{2}
\end{eqnarray}
Lagrangians (\ref{1}) and (\ref{2}) are equivalent on the
equations of motion for bosonic fields, from which it follows
that
\begin{equation}
\Delta^\delta\sim \bar q^C \epsilon^\delta\gamma^5 q,\quad
  \sigma\sim\bar qq,\quad
  \vec \pi\sim i\bar q\gamma^5\vec\tau q.
\end{equation}
However, it is more convenient to start our consideration from
Lagrangian (\ref{2}). Clearly, the $\sigma$ and $\vec\pi$ fields
are color singlets. Besides, the (bosonic) diquark field
$\Delta^{\delta}$ is a color antitriplet and a (isoscalar) singlet
under the chiral $SU(2)_L\times SU(2)_R$ group. Note further
that  $\sigma$ and $\Delta^\delta$ are (Lorentz) scalars, but
$\vec\pi$ are pseudoscalar fields. Hence, if $\vev{\sigma}\ne 0$,
then the chiral symmetry of the model is spontaneously broken
(at $m_o=0$), whereas $\vev{\Delta^\delta}\ne 0$ indicates  the
dynamical breaking of color symmetry. In the framework of
the Nambu--Gor'kov formalism
(for a recent relativistic treatment see, e.g. ref.~\cite{ohsaku})
quark fields are represented by  a bispinor
$\Psi=\left({q\atop q^C}\right)$. Integrating in the generating functional
based on the Lagrangian (\ref{2}) over
the quark fields, one obtains an effective  meson-diquark action of the
original model (1)
\begin{equation}
{\cal S}_{\rm
{eff}}(\sigma,\vec\pi,\Delta^\delta,\Delta^{*\delta})=-\int
d^4x\left[\frac{\sigma^2+\vec\pi^2}{4G_1}+
\frac{\Delta^\delta\Delta^{*\delta}}{4G_2}\right]-\frac i2{\rm
Tr}_{sfcx}\ln\left (\begin{array}{cc}
D^+ & K^-\\
K^+ & D^-
\end{array}\right ).
\label{9}
\end{equation}
Besides of an evident trace over the two-dimensional Nambu--%
Gor'kov matrix, the Tr-operation in (\ref{9}) stands for
calculating  the trace in spinor- ($s$), flavor- ($f$), color-
($c$) as well as four-dimensional coordinate- ($x$) spaces,
correspondingly. We have used also the following notations:
\begin{eqnarray}
 K^+=i\Delta^{*\delta}\epsilon^\delta\gamma^5,&\qquad&
 K^-=-i\Delta^{\delta}\epsilon^\delta\gamma^5,\nonumber\\
D^\pm=i\gamma^\nu\partial_\nu \pm\mu\gamma^0-m_o-\Sigma^\pm,&\qquad&
\Sigma^\pm=\sigma \pm i\gamma^5\vec\pi\vec\tau.
\label{10}
\end{eqnarray}
Let us introduce notations for the ground state expectation values of
the meson and diquark fields: $\vev{\sigma}=\sigma_o$,
$\vev{\vec\pi}=\vec\pi_o$,
$\vev{\Delta^{*\delta}}=\Delta^{*\delta}_o$,
$\vev{\Delta^{\delta}}=\Delta^{\delta}_o$ and
\[
(K^\pm_o, D^\pm_o, \Sigma^\pm_o)= (K^\pm, D^\pm,
\Sigma^\pm)~\bigg |_{~\sigma =\sigma_o, \vec\pi =\vec\pi_o,
\Delta^{\delta}=\Delta^{\delta}_o, ...}
\]
The quantities $\sigma_o,\pi^k_o, \Delta^\delta_o,
\Delta^{*\delta}_o$
($k,\delta=1,2,3$) correspond to the global minimum  of the effective
potential $V_{\rm {eff}}$ and can be found as a solution of the system of equations
\begin{eqnarray}
\frac{\partial V_{\rm {eff}}}{\partial\pi^k}=0,~~~~~
\frac{\partial V_{\rm {eff}}}{\partial\sigma}=0,~~~~~
\frac{\partial V_{\rm {eff}}}{\partial\Delta^\delta}=0,~~~~~
\frac{\partial V_{\rm {eff}}}{\partial\Delta^{*\delta}}=0,
\label{13}
\end{eqnarray}
usually called  gap equations. In (\ref{13}) we used the following
definition of the effective potential:
\begin{equation}
{\cal S}_{\rm {eff}}~\bigg |_{~\sigma,\vec\pi,\Delta^{\delta},\Delta^{\delta}=\rm {const}}
=- V_{\rm {eff}}(\sigma,\vec\pi,\Delta^{\delta},\Delta^{\delta})\int d^4x,
\label{130}
\end{equation}
where, in the spirit of the mean-field approximation, all fields are considered to be independent of $x$.
Without loosing any generality of
the consideration, one can search for a solution of  equations
(\ref{13})  in the form:
$(\Delta^1_o,\Delta^2_o,\Delta^3_o)$$\equiv (0,0,\Delta)$.
Moreover, we suppose also that $\vec\pi_o=0$. (The last
assumption is maintained by the observation that in the theory
of strong interactions P-parity is a conserved quantity.)
If $\Delta=0$, the color symmetry of the model remains intact. If
$\Delta\ne 0$, then $SU_c(3)$ symmetry is spontaneously broken
down to $SU_c(2)$, and the 2SC phase is realized in the model.
In this case, the system of gap equations (\ref{13})
is reduced to the following one:
\begin{eqnarray}
&&\frac{\sigma_o}{2G_1}=4iM\!\int\!\frac{d^4q}{(2\pi)^4E}
\left\{\frac{E^+}{q_0^2-(E^+)^2}+\frac{E^-}{q_0^2-(E^-)^2}+
\frac{2E^+}{D_+(q_0)}+
\frac{2E^-}{D_-(q_0)}\right\},\label{15}\\
&&\frac{\Delta}{4G_2}=4i\Delta\!\int\!\frac{d^4q}{(2\pi)^4}
\left\{\frac{1}{D_+(q_0)}+ \frac{1}{D_-(q_0)} \right\},
\label{16}
\end{eqnarray}
where $D_\pm(q_0)=q_0^2-(E^\pm)^2-|\Delta|^2$,
$E^\pm=E\pm\mu$, $E=\sqrt{\strut\vec q^{~\!2}+M^2}$, and
$M=m_o+\sigma_o$ is the constituent quark mass. In these and other
similar expressions, $q_0$ is a shorthand notation for
$q_0+i\varepsilon\cdot {\rm sgn}(q_0)$, where the limit
$\varepsilon\to 0_+$ must be taken at the end of all
calculations. This prescription correctly implements the role of
$\mu$ as chemical potential and preserves the causality
of the theory (see, e.~g. \cite{chodos}).

Let us now make a field shift $\sigma (x)\to\sigma (x)+\sigma_o$,
$\Delta^{*\delta}(x)\to\Delta^{*\delta}(x) +\Delta^{*\delta}_o$,
$\Delta^{\delta}(x)\to\Delta^{\delta}(x)+\Delta^{\delta}_o$ and then expand
the expression (\ref{9}) into a power series of the meson and
diquark fields. The second-order term ${\cal S}_{\rm
{eff}}^{(2)}$ of this expansion is responsible for the mass
spectrum of mesons and diquarks. It looks like
\begin{eqnarray}
 {\cal S}^{(2)}_{\rm
 {eff}}(\sigma,\vec\pi,\Delta^\delta,\Delta^{*\delta})=-\int\!d^4x
 \left[\frac{\sigma^2+\vec\pi^2}{4G_1}+
 \frac{\Delta^\delta\Delta^{*\delta}}{4G_2}\right]+&&\nonumber\\
+\frac i4{\rm Tr}_{sfcx}
\left\{S_0\left (\begin{array}{cc}
\Sigma^+, & -K^-\\
 -K^+, & \Sigma^-
\end{array}\right )\right.&&\left.\!\!\!\!\!\!\!\!S_0\left
(\begin{array}{cc}
\Sigma^+, & -K^-\\
 -K^+, & \Sigma^-
\end{array}\right )\right\},
  \label{11}
\end{eqnarray}
where $S_0$ is the quark propagator, which is an evident
2$\times$2 matrix in the Nambu--Gor'kov space
\begin{eqnarray}
S_0=   \left (\begin{array}{cc}
D^+_o & K^-_o\\
K^+_o &D^-_o
\end{array}\right )^{-1}\equiv  \left (\begin{array}{cc}
a & b\\
c & d
\end{array}\right ),
\label{12}
\end{eqnarray}
whose matrix elements can be found by means of the projection
operator technique \cite{kitai}:
\begin{eqnarray}
\label{121} &&a=\int\!\frac{d^4q}{(2\pi)^4}\,
e^{-iq(x-y)}\left\{
\frac{q_0-E^+}{q_0^2-(E^+)^2+|\Delta|^2\epsilon^3\epsilon^3}
\gamma^0\bar\Lambda_++
\frac{q_0+E^-}{q_0^2-(E^-)^2+|\Delta|^2\epsilon^3\epsilon^3}
\gamma^0\bar\Lambda_-\right\},\\
&&b=-i\Delta\epsilon^3\!\int\!\frac{d^4q}{(2\pi)^4}e^{-iq(x-y)}
\left\{\frac{1}{q_0^2-(E^+)^2+|\Delta|^2\epsilon^3
\epsilon^3}\gamma^5\bar\Lambda_-+
\frac{1}{q_0^2-(E^-)^2+|\Delta|^2\epsilon^3
\epsilon^3}\gamma^5\bar\Lambda_+\right\},\label{122}\\
&&c=i\Delta^*\epsilon^3\!\int\!\frac{d^4q}{(2\pi)^4}e^{-iq(x-y)}\left
\{\frac{1}{q_0^2-(E^+)^2+|\Delta|^2\epsilon^3
\epsilon^3}\gamma^5\bar\Lambda_++
\frac{1}{q_0^2-(E^-)^2+|\Delta|^2\epsilon^3
\epsilon^3}\gamma^5\bar\Lambda_-\right\},\label{123}\\
&&d=\int\!\frac{d^4q}{(2\pi)^4}e^{-iq(x-y)}\left\{
\frac{q_0+E^+}{q_0^2-(E^+)^2+|\Delta|^2\epsilon^3\epsilon^3}
\gamma^0\bar\Lambda_-+
\frac{q_0-E^-}{q_0^2-(E^-)^2+|\Delta|^2\epsilon^3\epsilon^3}
\gamma^0\bar\Lambda_+\right\},\label{124}
\end{eqnarray}
where $\bar\Lambda_\pm=\frac 12 \left
(1\pm\frac{\gamma^0(\vec\gamma\vec p-M)}{E}\right )$ are
projectors on the solutions of the Dirac equation with
positive/negative energy,
$\epsilon^3\epsilon^3={\rm diag}(-1,-1,0)$ is the projector
(up to a sign) in the color space. Quantities
(\ref{121})--(\ref{124}) are nontrivial operators in the
coordinate, spinor, and color spaces, but they are
unit operators in flavor space.

A tedious but straightforward analysis of ${\cal S}_{\rm
{eff}}^{(2)}$ shows that  there arises a mixing between
the $\sigma$ and $\Delta^3$ fields. We have
\begin{eqnarray}
&& {\cal S}^{(2)}_{\rm
 {eff}}(\sigma,\vec\pi,\Delta^\delta,\Delta^{*\delta})=
 {\cal S}^{(2)}_{\pi\pi}(\vec\pi)+
 {\cal S}^{(2)}_{\sigma 3}(\sigma,\Delta^3,\Delta^{*3})+{\cal
 S}^{(2)}_{1}(\Delta^{1},\Delta^{*1})+{\cal
 S}^{(2)}_{2}(\Delta^{2},\Delta^{*2}),
\label{20}
\end{eqnarray}
where ($\rho=1,2$)
\begin{eqnarray}
&& {\cal S}^{(2)}_{\rho}(\Delta^{\rho},\Delta^{*\rho})=
-\!\int\!d^4x\frac{\Delta^\rho\Delta^{*\rho}}{4G_2} +\frac i2{\rm
Tr}_{sfcx}
\left\{a\Delta^\rho\epsilon^\rho\gamma^5d\Delta^{*\rho}\epsilon^\rho
\gamma^5\right\}. \label{21}
\end{eqnarray}
In the following we are going to search for gapless bosonic
excitations of the 2SC quark-matter ground state of the model only.
Since we are mainly interested in the diquark sector, the expression
for the $\pi$-meson effective action ${\cal
S}^{(2)}_{\pi\pi}(\vec\pi)$ is
omitted here. Moreover, ${\cal S}^{(2)}_{\sigma 3}(\sigma,\Delta^3,
\Delta^{*3})$ has a rather cumbersome form, so we also do not show
it. (If needed, both these expressions can be obtained
immediately from (\ref{11}).)

Before proceeding further, we should fix the
parameters of our model. Here, we choose the parametrization
procedure given in \cite{volk}. We simplify that model by ignoring
the corrections to the pion-quark coupling constant that come
from  the mixing between the pion and axial-vector. To fix the
model parameters, we require the model to reproduce the
Goldberger-Treiman relation $M g_\pi=F_\pi$, where $F_\pi\approx
93 $ MeV is the pion weak-decay constant, and $g_\pi$ is the
pion-quark coupling constant determined in the local NJL model
as follows
\begin{equation}
g_{\pi}^{-2}=\frac{-i
N_c}{(2\pi)^4}\!\int\!\frac{d^4q}{(q^2-M^2)^2}.
\label{coupl}
\end{equation}
(This integral is divergent, and a regularization is supposed to be
implemented here. In our work we use a three-dimensional cutoff to
make such integrals meaningful.)

Then, we require the observed pion mass ($\sim 140$ MeV) to be
reproduced by the local NJL model with our parameters.
After this, one more condition is needed to complete the parameter
fixing procedure. We consider two possibilities to finalize it:
i) fix the model parameters so that the QCD sum rules
estimate for the chiral condensate $\vev{\bar qq}\approx(-245$~MeV$)^3$
is reproduced in our model, ii) choose the model parameters in a way
that allows one to obtain the correct description of the
$\rho\to\pi\pi$ decay, as it was done in \cite{volk}.
In the first case (let us refer to this set of parameters as set-A),
the cutoff $\Lambda$ is about 618 MeV, the four-quark interaction
constant $G_1$ is equal  to 5.86 GeV$^{-2}$. The remaining constant,
$G_2$, is chosen according to the Fierz-transformation as
$G_2=3G_1/4= 4.40$ GeV $^{-2}$. The current quark mass $m_o$ is fixed
via the gap equations to the value: $5.7$ MeV. The constituent quark
mass in the vacuum (at zeroth temperature and chemical potential)
amounts to 350 MeV.

In the second case, we obtain a different set of model parameters,
which will be referred to, throughout the paper, as the  set-B.
Imposing the condition that an extension of this model to
the vector channel would give the experimental value for the
$\rho$-meson width  (the $\rho$-meson decays
mostly into a couple of charged pions, and the decay is described
by the constant $g_\rho\approx 6.1$.) and keeping in mind that in the
local NJL model one gets a natural connection between $g_\rho$ and
$g_\pi$ (discarding pion--axial-vector transitions),
$\sqrt{6}g_\rho=g_\pi$, we obtain $\Lambda=856$ MeV, $G_1=2.49$
GeV$^{-2}$, $G_2=3G_1/4=1.87$ GeV$^{-2}$, and $m_0=3.6$ MeV.
In the vacuum the constituent quark mass is 233 MeV, which is small,
compared to the case, where the chiral condensate is used to fix the
model parameters\footnote{A larger
quark mass can be obtained if one takes into account the mixing
between the pion and axial-vector. However, we omit it in our
paper for simplicity and for the reason that this mixing is
negligible in the 2SC phase.}.
A similar procedure has been
implemented in \cite{bvy03} in the chiral limit. Here, we keep
the current quark mass non-zero in order to reproduce the pion
mass in the vacuum.

\begin{figure}
\begin{center}
\includegraphics[scale=0.5]{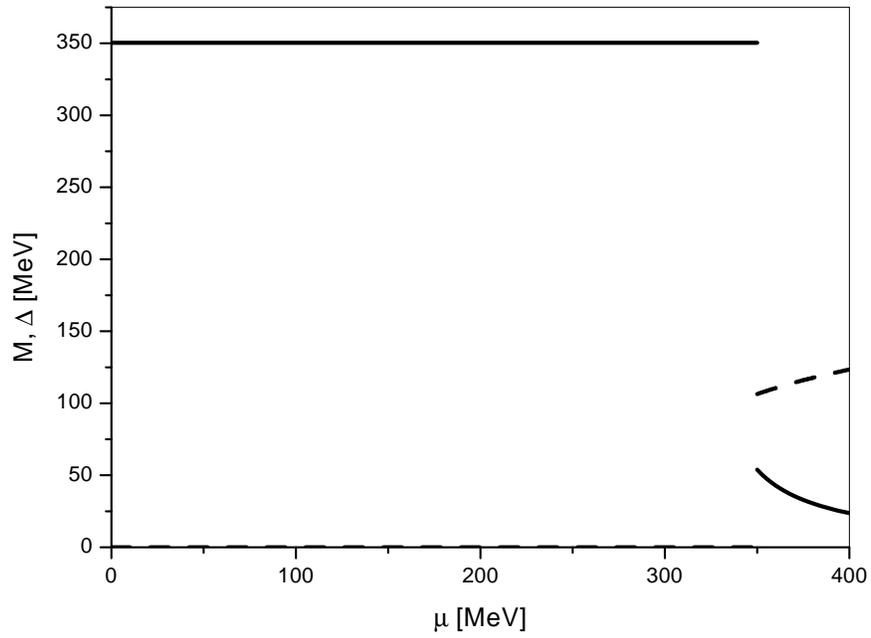}
\end{center}
\caption{The constituent quark mass $M$ (solid line) and the
color gap $\Delta$ (dashed line) as  functions of the chemical
potential (parameter set-A). } \label{MDeltamu1}
\end{figure}

\begin{figure}
\begin{center}
\includegraphics[scale=0.5]{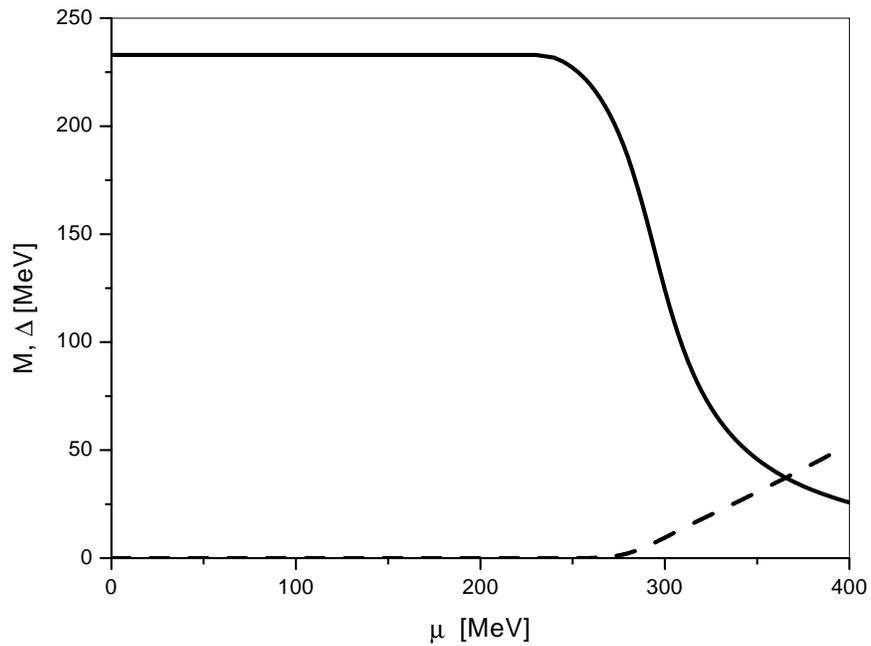}
\end{center}
\caption{The constituent quark mass $M$ (solid line) and the
color gap $\Delta$ (dashed line) as  functions of the chemical
potential (parameter set-B). } \label{MDeltamu2}
\end{figure}

Solving the gap equations at a fixed chemical potential, we obtain
the constituent quark mass $M$ and the color gap $\Delta$ that
satisfy the global minimum of the effective potential. In
Figs.~\ref{MDeltamu1} and \ref{MDeltamu2}, one can see the quark mass
$M$ and the color gap $\Delta$ \textit{vs.} the chemical potential
for parameter sets A and B.
For the parameter set A, one observes a first-order phase
transition from hadron matter to the 2SC phase. The gaps (constituent
quark mass and color gap) change abruptly when the chemical potential
reaches the vacuum value of $M$ (see Fig.~\ref{MDeltamu1}). On the
contrary, if the parameter set-B is chosen, the gaps  reveal a rather
smooth character near the phase transition, which is typical for a
phase transition of the second-order (see Fig.~\ref{MDeltamu2}).

\section{Gapless excitations in the diquark sectors}

\begin{figure}
\begin{center}
\includegraphics[scale=0.5]{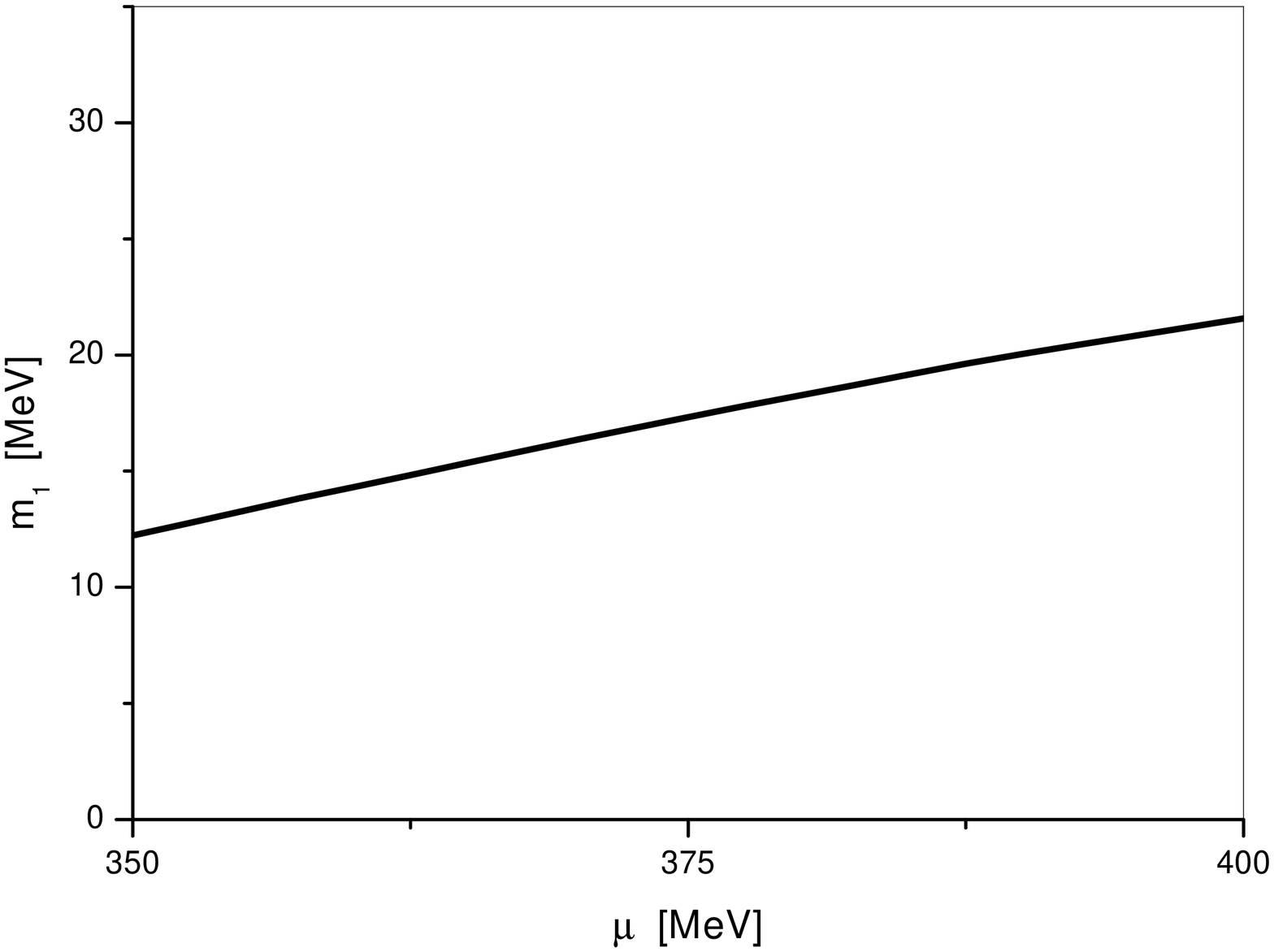}
\end{center}
\caption{The diquark mass $m_1=H(0)/H^\prime(0)$ as a function of the
chemical potential (parameter set-A).} \label{m1plot1}
\end{figure}

\begin{figure}
\begin{center}
\includegraphics[scale=0.5]{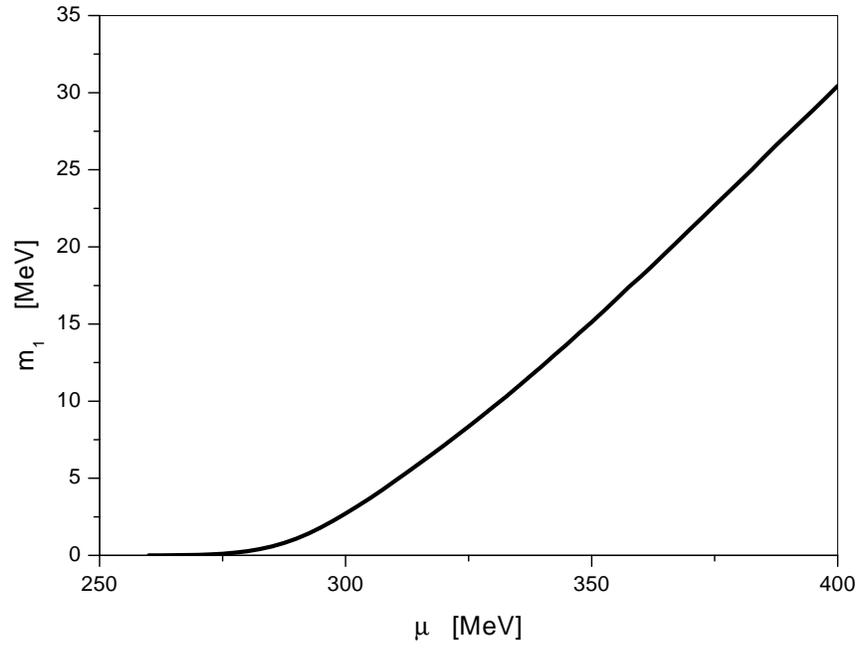}
\end{center}
\caption{The diquark mass $m_1=H(0)/H^\prime(0)$ as a function of the
chemical potential (parameter set-B).} \label{m1plot2}
\end{figure}

Since  $\Delta^1$ in (\ref{20}) is not mixed with other fields,
let us, first of all, focus on the diquark $\Delta^1$-sector. If
$\Delta^1(x)=(\varphi_1 (x)+i\varphi_2 (x))/\sqrt{2}$,
$\Delta^{*1}(x)=(\varphi_1 (x)-i\varphi_2 (x))/\sqrt{2}$, then
the effective action ${\cal S}^{(2)}_{1}(\Delta^{1},\Delta^{*1})$
from (\ref{20}) can be presented in the form:
\begin{eqnarray}
&&{\cal S}^{(2)}_{1}(\Delta^{1},\Delta^{*1})\equiv {\cal
S}^{(2)}_{1}(\varphi_1, \varphi_2)= -\frac 12\! \int\! d^4x\,
d^4y\varphi_k (x) \Gamma_{kl}(x-y)\varphi_l(y), \label{230}
\end{eqnarray}
where the $2\times 2$ matrix $\Gamma(x-y)$ is the inverse propagator
of the $\varphi_1, \varphi_2$-fields (summation over $k,l=1,2$ is
implied in (\ref{230})). Its matrix elements can be
found via a second variation of ${\cal S}^{(2)}_{1}$:
\begin{eqnarray}
&& \Gamma_{kl}(x-y)=-\frac{\delta^2{\cal S}^{(2)}_{1}}
{\delta\varphi_l(y)\delta\varphi_{k}(x)}.
\label{23a}
\end{eqnarray}
In  momentum space, the Fourier-transformed components of the
$\Gamma$-matrix have the following structure
\begin{eqnarray}
\overline{\Gamma_{11}}(p)=\overline{\Gamma_{22}}(p)=\frac 12\left
(\overline{\Gamma_{\Delta^*\Delta}}(p)+
\overline{\Gamma_{\Delta^*\Delta}}(-p)\right ),~~~
\overline{\Gamma_{12}}(p)=-\overline{\Gamma_{21}}(p)=\frac i2\left
(\overline{\Gamma_{\Delta^*\Delta}}(p)-
\overline{\Gamma_{\Delta^*\Delta}}(-p)\right ).
\label{23c}
\end{eqnarray}
The derivation of these relations is given in Appendix, where the
quantity $\overline{\Gamma_{\Delta^*\Delta}}(p)$ is also
presented (see eq.~(\ref{250})). The particle
dispersion laws are defined by zeros of
det$\overline{\Gamma}(p)$ in the $p_0$-plane, i.~e. by the
equation:
\begin{eqnarray}
{\rm det}\overline{\Gamma}(p)=\overline{\Gamma_{11}}(p)
\overline{\Gamma_{22}}(p)-\overline{\Gamma_{12}}(p)
\overline{\Gamma_{21}}(p)=
\overline{\Gamma_{\Delta^*\Delta}}(p)
\overline{\Gamma_{\Delta^*\Delta}}(-p)=0.
\label{26b}
\end{eqnarray}
Clearly, the solutions of (\ref{26b}) with positive/negative $p_0$
correspond to particles/antiparticles. Let us first
put $p=(p_0,0,0,0)$. In this case, det$\overline{\Gamma}(p_0)$
is an even function of $p_0$, so the solutions of eq.~(\ref{26b})
in the $p_0^2$-plane might be searched for. They are
the squared masses of excited states in the $\Delta^1$-sector
of the model. Moreover, all these formulae are then greatly
simplified. Indeed, it follows from (\ref{250}) at
$\vec{p}^{~2}=0$ that
\begin{eqnarray}
&&\overline{\Gamma_{\Delta^*\Delta}}(p_0)=4ip_0\!\int\!\frac{d^4q}
{(2\pi)^4}
\left\{\frac{1}{(p_0+q_0+E^+)D_+(q_0)}+\frac{1}{(p_0+q_0-E^-)D_-(
q_0)}\right\}\equiv -4p_0H(p_0), \label{26}
\end{eqnarray}
where it is possible to integrate over $q_0$, using the
following prescription: $(p_0+q_0)\to
(p_0+q_0)+i\varepsilon\cdot {\rm sgn}(p_0+q_0)$ and $q_0\to
q_0+i\varepsilon\cdot {\rm sgn}(q_0)$ with $\varepsilon\to
0_+$ (see also comments after formula (\ref{16})), and
\begin{eqnarray}
H(p_0)=-\frac 12\int\!\frac{d^3q} {(2\pi)^3}
\left\{\frac{1}{(p_0+E^++E_\Delta^+)E_\Delta^+}+
\frac{\theta(E^-)}{(p_0-E^--E_\Delta^-)E_\Delta^-}+
\frac{\theta(-E^-)}{(p_0-E^-+E_\Delta^-)E_\Delta^-}\right\}.
\label{26c}
\end{eqnarray}
In (\ref{26c}) $E_\Delta^\pm=\sqrt{(E^\pm)^2+|\Delta|^2}$.
Since the integral in the RHS of this equation is ultraviolet
divergent, we regularize it, as the other divergent integrals,
by using a three-dimensional cutoff  $\Lambda$, i.~e.{}
$\vec q^{~2}\leq \Lambda^2$. Then,  eq.~(\ref{26b}) is
transformed to
\begin{equation}
p_0^2H(p_0)H(-p_0)=0,
\label{260b}
\end{equation}
from which it is evident that there is a solution $p_0^2=0$,
corresponding to a gapless excitation of the 2SC ground state.
Since the function $H(p_0)$ does not vanish at $p_0=0$ and
$\mu\ne 0$, this solution may be identified in the case of finite $\vec p^2$ with a
NG-boson with the quadratic dispersion
law: $p_0\sim \vec p^{~\!2}/H(0)$ (see Appendix). Now, let us suppose that at some
nonzero value $p_0=-m_1$ the function $H(p_0)$ has a zero, i.~e.
$H(-m_1)=0$. Then, ${\rm det}\Gamma (p_0)$ has two zeroes $p_0=\pm
m_1$, i.~e. the point $p_0^2=m_1^2$ is a solution of
eq.~(\ref{260b}), and the second bosonic excitation  of this
sector has the nonzero mass $m_1$. In the low-$p_0$ expansion we
have: $H(p_0)=H(0)+p_0H^\prime(0)+\cdots$, where
\begin{eqnarray}
&&H(0)=\frac 12\int\frac{d^3q}{(2\pi)^3}
\left\{\frac{\theta(E^-)}{(E^-+E_\Delta^-)E_\Delta^-}+
\frac{\theta(-E^-)}{(E^--E_\Delta^-)E_\Delta^-}-
\frac{1}{(E^++E_\Delta^+)E_\Delta^+}\right\},\nonumber\\
&&H'(0)=\frac 12\int\frac{d^3q}{(2\pi)^3}
\left\{\frac{\theta(E^-)}{E_\Delta^-(E^-+E_\Delta^-)^2}+
\frac{\theta(-E^-)}{(E^--E_\Delta^-)^2E_\Delta^-}+
\frac{1}{E_\Delta^+(E^++E_\Delta^+)^2}\right\}.
\label{26d}
\end{eqnarray}
Thus, in this approximation $m_1=H(0)/H^\prime(0)$. The $m_1$
\textit{vs.} $\mu$ plots for particular values of the coupling
constants $G_1, G_2$ and the cutoff parameter $\Lambda$ discussed at
the end of section 2 (or, equivalently, for the values $\Delta$ and $M$
from Figs.~\ref{MDeltamu1} and \ref{MDeltamu2}) are presented in
Figs.~\ref{m1plot1} and \ref{m1plot2}. In Fig.~\ref{m1plot1} the
values of $m_1$ are calculated for $\mu=350\div400$ MeV when the
parameter set-A is involved. The leftmost point corresponds to the
condition of the phase transition ($\mu\approx 350$ MeV).
At smaller $\mu$ the color gap equals zero and eqs.~(\ref{26d})
cannot be applied, since they have been derived by using
eq.~(\ref{16}) with $\Delta\not=0$. Similarly, in Fig.~\ref{m1plot2}
the axis $\mu$ begins with $250$ MeV. At smaller $\mu$, (but grater
than 233 MeV), the value of $m_1$ is almost zero. Common to both
parameter sets A and B is that the values of $m_1$ are small
and thus justify the series expansion of the function $H(p_0)$ around
$p_0=0$.%
\footnote{The numerical investigation
shows that the mass value $m_1$ found in the low-$p_0$ approximation (see
Figs.~\ref{m1plot1} and \ref{m1plot2})
differs from the exact solution of the equation $H(m_1)=0$ by no more
than 10$\%$.}
Analogous results are obtained for the $\Delta^2$-sector of the
model. There, as in the $\Delta^1$-sector,  only one massless
boson with the quadratic dispersion law as well as a massive one
with the mass $m_2\equiv m_1$ are found. Evidently, the two massless
and two massive bosons in the $\Delta^1$, $\Delta^2$-sectors form
antidoublets with respect to the unbroken $SU_c(2)$ symmetry.

Now let us consider the sector with mixing of the complex
diquark $\Delta^3$-field and the scalar $\sigma$-meson.
Introducing new real fields: $\phi (x)=(\Delta^3(x)+
\Delta^{*3}(x))/\sqrt{2}$, $\psi (x)=i(\Delta^{*3}(x)
-\Delta^3(x))/ \sqrt{2}$, we rewrite the  effective action
${\cal S}^{(2)}_{\sigma 3}$ (here we omit the cumbersome
expression for this quantity, however it can be easily reproduced
from (\ref{11})) in the following form:
\begin{eqnarray}
&& {\cal S}^{(2)}_{\sigma 3}(\sigma,\Delta^{3},\Delta^{*3})=
-\frac12\!\int\!d^4x\,d^4y(\sigma (x),\phi (x),\psi (x))~\Pi
(x-y)~( \sigma (y),\phi (y),\psi (y))^t, \label{270}
\end{eqnarray}
where $\Pi (x-y)$ is the inverse propagator of the
$\sigma$-meson and the diquark fields $\phi,\psi$.
Clearly, it is a 3$\times $3 matrix, whose elements can be obtained
from (\ref{270}) by taking all second variational derivatives of
${\cal S}^{(2)}_{\sigma 3}$ over the fields $\sigma,\phi$, and
$\psi$. We have calculated the momentum-space matrix
$\overline{\Pi}(p)$ (corresponding explicit formulae and
relations are omitted), and then, putting $p=(p_0,0,0,0)$, found
that det$(\overline{\Pi}(p_0))$ has only one zero at $p_0^2=0$.
Therefore, only one NG-boson, which is an $SU_c(2)$-singlet,
exists in this sector (as in the
$\Delta^1,\Delta^2$-sectors, each zero of
det$(\overline{\Pi}(p_0))$ in the $p_0^2$-plane means the mass
squared of the ground-state excitation, this time in the mixed
$\sigma$-$\Delta^3$ sector). \footnote{A detailed investigation of
the meson/diquark spectrum including the $\sigma$-$\Delta^3$ mixing is
in preparation.}

As it was mentioned above, in the 2SC phase of the model, where
$\Delta\ne 0$,  the color $SU_c(3)$  symmetry of the ground state is
spontaneously broken down to the $SU_c(2)$ group. Hence, in
accordance with the Goldstone theorem, it is generally expected that
five NG-bosons would appear in the theory. However, we have
proved that in the $\Delta^1,\Delta^2,\Delta^3$-sectors of the
model only three massless bosons exist, which can be identified with
NG-bosons. Therefore, there seems to be
a deficiency of two NG-bosons in the framework of the model under
consideration. In spite of this fact, there are, however, no
contradictions with the NC theorem, since two of the three found
NG-bosons have quadratic dispersion laws for $|\vec p|\to
0$. In addition, the model contains also light massive diquarks.

Usually, in the framework of the NJL model, the 2SC phase
is studied in the region of coupling constants (\ref{1a}). In this
case, at $\mu =0$ we have a chirally noninvariant phase even if
$m_o\to 0$, and the transition to the 2SC phase occurs at some
finite value of the chemical potential. Formally, however, one
can consider the region
\begin{equation}
\omega =\{(G_1,G_2):G_2>\pi^2/(4\Lambda^2),~~
\pi^2(G_2-G_1)>2G_1G_2\Lambda^2\},
\label{omega}
\end{equation}
where (even at $\mu =0$) the 2SC phase is realized \cite{klim2}.
As can easily be seen from (\ref{26d}), in the case of $\mu =0$
we have $H(0)=0$ and $m_1=m_2=0$, and the total number of
gapless excitations (NG-bosons) in the $\Delta^1$- and
$\Delta^2$-sectors of the model equals four. They form two massless
$SU_c(2)$-antidoublets. Apart from this, there is one NG-boson, which
is an $SU_c(2)$-singlet, in the mixed $\sigma$-$\Delta^3$ sector.
Naturally, all these excitations have a linear dispersion law, as it
should be in the relativistically invariant case (see Appendix).
However, at an arbitrary small $\mu$, particles from one of these
antidoublets acquire nonzero masses, and the dispersion laws for
the two massless particles from the remaining antidoublet are changed
essentially from the linear to quadratic laws.

Note finally, that in the above consideration we used the
real/imaginary part parametrization for diquark fields:
\begin{equation}
\Delta^1(x)=(\varphi_1 (x)+i\varphi_2 (x))/\sqrt{2},~~~
\Delta^2(x)=(\tilde\varphi_1 (x)+i\tilde\varphi_2 (x))/\sqrt{2},~~~
\Delta^3(x)-\Delta=(\phi (x)+i\psi (x))/\sqrt{2},
\label{I}
\end{equation}
where $\Delta^\rho(x)$ are the unshifted diquark fields from
Lagrangian (\ref{2}) and $\Delta=\vev{\Delta^3}$.
However, one can use also an alternative "polar coordinate"
parametrization:
\begin{equation}
\vect\Delta^1(x)\\ \Delta^2(x)\\ \Delta^3(x)\evect =
\exp\left\{-i\sum_a\frac{
\lambda_a^t\xi_a(x)}{\sqrt{2}\Delta}\right\}
\frac 1{\sqrt{2}}\vect0\\0\\\sqrt{2}\Delta +\eta (x)\evect =
\frac 1{\sqrt{2}}\vect \xi_5-i\xi_4 \\ \xi_7-i\xi_6 \\ \sqrt{2}\Delta
+\eta +i\frac 2{\sqrt{3}}\xi_8\evect +o(\xi_i,\eta),
\label{II}
\end{equation}
where $\lambda_a$ are Gell-Mann matrices and the summation over
$a=4,..,8$ is implied (the form of an exponential multiplier in
(\ref{II}) corresponds to the fact that $\Delta^\rho(x)$ is an
$SU_c(3)$-antitriplet). Comparing (\ref{I}) and (\ref{II}) at small
$\xi_i,\eta$, we see that
\begin{equation}
\varphi_1 =\xi_5,~~\varphi_2=-\xi_4,~~
\tilde\varphi_1=\xi_7,~~\tilde\varphi_2=-\xi_6,~~
\phi=\eta,~~\psi=\frac 2{\sqrt{3}}\xi_8.
\label{III}
\end{equation}
Further, one should insert (\ref{II}) into (\ref{9}) and expand the
resulting expression into a series of meson- and $\xi_a,\eta$ fields.
It is easily seen that in the second order of the new variables the
effective action is a sum similar to (\ref{20}). In particular, it
means that the diquark fields $\xi_4,\xi_5$ are decoupled from other
fields, and their effective action looks like (\ref{230}) with the
replacement (\ref{III}) used. So, some elements of the inverse
propagator matrix for $\xi_4,\xi_5$ fields differ by a sign from
$\Gamma_{kl}(x-y)$ (\ref{23a}), but the determinant of the whole
momentum space matrix is not changed. Hence, the mass spectrum, the
dispersion laws etc., in the $\xi_4,\xi_5$-fields are the same,
as in the case with old fields $\varphi_1,\varphi_2$. Similar
conclusions are valid for the rest of new variables, i.~e. such
physical characteristics of the model, as particle masses,
dispersion laws etc., do not depend on the choice of field
parametrizations, (\ref{I}) or (\ref{II}).
In particular, the number of abnormal NG-bosons, which equals three
in the model under consideration, is a parametrization invariant
quantity\footnote{In papers \cite{ms,sss} an abnormal number of
NG-bosons in the same $SU(2)\times U(1)$ toy model was discovered.
But in \cite{ms} the parametrization (\ref{I}) was used for
the scalar doublet, whereas in \cite{sss} a "polar coordinate"
parametrization, similar to (\ref{II}), was used. Evidently, their
results coincide, i.e. the abnormal number of NG-bosons, particle
masses, formulae for dispersion laws are the same.}.

\section{Color--asymmetry and masses of scalar diquarks}

As it was noted in \cite{rebhan}, the color superconducting phase
is  automatically color-neutral in QCD, but this does not hold for
the NJL model under consideration. Massive scalar diquarks and, as a consequence,
the abnormal number of NG-bosons in the model discussed in previous sections testify for this.
As mentioned in the Introduction, for the number of NG-bosons to equal the number of broken
generators of the underlying symmetry group (in our case $SU_c(3)$),
the ground-state expectation value  for the commutator of any two
broken symmetry generators $Q_a$ and $Q_b$
must be zero: $\vev{[Q_a,Q_b]}=0$ \cite{sss}. This criterion is not fulfilled  for
the NJL model considered here, namely, the ground state expectation value of the color charge operator
$Q_8=\bar q\gamma^0\lambda^8q$ does not vanish, and, therefore, the usual counting rule for
NG-bosons is not applicable.

Let us look at the function $H(p_0)$ at $p_0=0$. One can rewrite the expression
(\ref{26d}) in a different form:
\begin{equation}\label{H0}
H(0)=\frac12\int\frac{d^3q}{(2\pi)^3}\frac{1}{\Delta^2}
 \left(-2\theta(\mu-E)+\frac{E^+}{E^+_\Delta}-\frac{E^-}{E^-_\Delta} \right).
\end{equation}
This formula can be expressed in terms of quark densities.
For this purpose, we need the expression of  the thermodynamical potential of the system,
which in the mean-field approximation
is equal to the value of $V_{\rm {eff}}$ in its global minimum point, supplied by the
gap equations, i.~e.
\begin{equation}
\Omega(\mu)=V_{\rm {eff}}(\sigma_o,\vec\pi_o,
\Delta^{\delta}_o,\Delta^{*\delta}_o),
\end{equation}
where $V_{\rm {eff}}$ was given in (\ref{130}). Further,
one can divide  the thermodynamical potential into several terms:
\begin{equation}
\Omega(\mu)=\sum\limits_{c=1}^{3}\Omega_c(\mu)+\frac{(M-m_o)^2}{4G_1}+\frac{|\Delta|^2}{4G_2},
\end{equation}
where $\Omega_c(\mu)$ is the contribution from the color-$c$ quark,
defined as follows:
\begin{eqnarray}
\Omega_1(\mu)=\Omega_2(\mu)&=&-2\int\frac{d^3q}{(2\pi)^3}
\left(E^+_\Delta+E^-_\Delta\right), \nonumber\\
\Omega_3(\mu)&=&-2\int\frac{d^3q}{(2\pi)^3}\left(|E^+|+|E^-|\right).
\label{Omegac}
\end{eqnarray}
The contribution from color-1 and color-2 quarks, $\Omega_{1(2)}(\mu)$ are equal
due to the remaining $SU_c(2)$ color symmetry after the global $SU_c(3)$ symmetry
group is spontaneously broken in the color superconducting phase. The
third color quark, which does not condensate,  gives $\Omega_3(\mu)$.
Then, one can calculate the densities $n_c$ of color-1, -2, and -3 quarks separately without
introducing additional chemical potentials for each color simply by
differentiating $\Omega_c(\mu)$ over $\mu$, i.~e.
\begin{equation}
n_c=-\frac{\partial\Omega_c(\mu)}{\partial\mu}, \qquad c=1,2,3;
\end{equation}
\begin{equation}\label{nc}
n_1=n_2=2\int\frac{d^3q}{(2\pi)^3}
\left(\frac{E^+}{E^+_\Delta}-\frac{E^-}{E^-_\Delta} \right), \qquad
n_3=4\int\frac{d^3q}{(2\pi)^3}\theta(\mu-E).
\end{equation}
Eq.~(\ref{H0}) can thus be transformed to
\begin{equation}\label{H0b}
H(0)=\frac{1}{8\Delta^2}(n_1+n_2-2n_3)=
\frac{\sqrt{3}\vev{Q_8}}{8\Delta^2}.
\end{equation}

From eq.~(\ref{H0b}) and the fact that the diquark mass $m_1= H(0)/H'(0)$ is
found to be nonzero (see Figs. 3 and 4), one concludes that the 2SC-ground state
in the NJL model under consideration has finite color charge $\vev{Q_8}$.
Evidently, this fact is related to the inequality of color quark densities
$n_{1,2}>n_3$ for paired (1,2) and unpaired (3) quarks, i.~e. to
color-asymmetry. Moreover,
as discussed in Appendix, the nonzero value of $H(0)$ is responsible for the
abnormal dispersion law $p_0\sim \vec p^{\;2}/H(0)$ for two NG-bosons.

To restore the
local color neutrality in the NJL model, one can, e.~g., include an additional
term, $\mu_8 Q_8$, and impose the color neutrality condition:
\begin{equation}
\vev{Q_8}=-\frac{\partial \Omega(\mu,\mu_8)}{\partial \mu_8}=0,
\end{equation}
or consider the superconducting matter as composed of colored domains,
with the global color charge being equal zero. The latter, however,
goes beyond the mean-field approximation exploited here and
should be considered as an approximation for the case of large
colored domains. Anyway, the true ground state
should be checked to give the absolute minimum
of the thermodynamical potential.

The inclusion of an additional chemical potential $\mu_8$ ``by hand'', however,
does not contradict QCD. As discussed in \cite{rebhan},
a non-vanishing expectation value of the gluon field $A_0^8$, corresponding to the
eighth generator in the $SU_c(3)$ group, gives rise to a non-vanishing
value of $\mu_8$, which then cancels certain tadpole contributions and
restores the color neutrality. This cannot occur in the NJL model automatically because
there are no gluons in it.

\section{Summary and discussions}

Recently, it has been shown that in
QCD with a nonzero strangeness chemical potential the number of
NG-bosons can be less than the number of broken symmetry generators \cite{ms,sss}, so
that the usual counting rule for NG-bosons evidently does not hold.
In the present paper, we have presented another relativistic model providing us with
an abnormal number of NG-bosons. This is an NJL--type model with two
light-quark flavors, where a single quark
chemical potential is taken into account. In the color
superconducting phase of this model,
the color $SU_c(3)$ group is spontaneously broken down to
$SU_c(2)$, i.~e. the number of broken symmetry generators
equals five. Despite of this fact,
there appeared only three NG-bosons in the 2SC
phase; two of them form an antidoublet with respect to the unbroken
$SU_c(2)$ subgroup and have quadratic
dispersion laws. The remaining one is an $SU_c(2)$-singlet, so there
are no contradictions with the NC theorem \cite{nielsen}. Moreover,
there exists an
$SU_c(2)$-antidoublet of light diquarks $\Delta^1,\Delta^2$ with masses $m_1$, the behavior of
which at final chemical potential $\mu$ is shown in
Figs.~3 and 4 for the parameter sets A and B, respectively.

Most interestingly,  we found that the diquark masses $m_1=H(0)/H'(0)$ are proportional to
the ground state average of the color charge operator $\vev{Q_8}=1/\sqrt3(n_1+n_2-2n_3)$.
Here, the quark densities $n_1$, $n_2$ of paired quarks with color-1,2 are
equal due to the remaining $SU_c(2)$
symmetry and larger than the density $n_3$ of the unpaired color-3 quark.
Thus, the ground state of
the considered NJL-type model is evidently color-asymmetric. It is just the appearance of this
nonvanishing expectation value $\vev{Q_8}$ which simultaneously leads to a violation of the criterion
for the equality of the number of NG-bosons and broken symmetry generators \cite{sss}, and
to the appearance of the quadratic dispersion law for the NG-diquarks (compare eqs.~(38), (A6)).

In the investigations of
color superconductivity in the two-flavored QCD, five scalar NG-bosons are argued to exist.
They are mixing with gluons and required as longitudinal components
for five massive gluons (color Meissner effect) \cite{msw}.
The fact that we find an abnormal number of three NG-bosons
does not directly contradict QCD.
Namely, in the color superconducting phase of QCD a nonvanishing condensate
of the eighth gluon field component $A^8_0$ appears, which induces a color chemical potential
$\mu_8$ and cancels certain tadpole contributions,
responsible for the nonvanishing color charge of the ground state \cite{rebhan}.
This mechanism just leads to color neutrality.
Obviously, there arises the question, how can one reconcile
the considered NJL approach
with QCD?
Insofar as NJL-type  models do not contain gluon fields, the required contribution
from the condensed eighth gluon field does not follow automatically. Therefore, the
tadpole contribution becomes unbalanced by gluon contributions. It can be
be cancelled ``by hand'', e.~g.{} with the help of the additional color chemical
potential $\mu_8$ simulating the omitted gluon contribution. In such enlarged NJL models
the condition of color neutrality of the ground state, i.~e.{} $\vev{Q_8}=0$,
can be imposed as an additional physical
requirement \cite{huang}. According to our results, one gets in this case two additional
NG-bosons as well as a change of quadratic dispersion laws into normal linear ones. Thus,
the abnormal number of three NG-bosons might, in principle, become converted into the
normal number five found in QCD.

Nevertheless, what to do now with these five NG-bosons arising  in the color neutral 2SC-phase of
an enlarged NJL-type model? Clearly, in the framework of the standard NJL model one cannot apply
the color Meissner effect, in order to absorb these NG-bosons into
longitudinal degrees of freedom of massive gluons.
A possible way out of this problem could be an extension of the NJL mode via
the inclusion of perturbative gluons, fluctuating
around the original low-energy gluonic fields (realized e.~g.{} by instantons or other non-perturbative
background fields, which were already integrated out to yield the effective four-quark interactions)
\cite{Diakonov}. Obviously, the color Meissner effect could then be realized for such
perturbative gluons.

In addition, let us remark that the above discussions refer to the case of
locally neutral quark matter, where color neutrality should indeed
be required. On the other hand, global color
neutrality does not demand local color neutrality:
for example, one can assume the ground state as being composed of
colored domains. Anyway, to decide which of the considered ground states
is in favor, the absolute minimum of the thermodynamic potential should
be calculated in each case. (Note that the presence of domains means
a violation of translational invariance and is not consistent with
the simple mean-field approach.)

Finally, notice that in the framework of the considered NJL-type of model there
arises a mixing between the $\sigma$-meson and $\Delta^3$-diquark in
the 2SC phase. The behavior of $\sigma$-, $\Delta^3$- as well as
$\pi$-meson masses {\it vs.} $\mu$ in the 2SC phase
is now under consideration.

\section*{Acknowledgments}

This work has been supported in part by DFG-project 436 RUS
113/477/0-2, RFBR grant 02-02-16194, the
Heisenberg--Landau Program, and the ``Dynasty'' Foundation.

\appendix
\section{The $\Gamma$-matrix elements and dispersion laws}

Using the effective action (\ref{21}), let us introduce auxiliary
quantities
\begin{eqnarray}
&& \Gamma_{\Delta^*\Delta}(z)=-\frac{\delta^2{\cal S}^{(2)}_{1}}
{\delta\Delta^1(y)\delta\Delta^{*1}(x)}=\frac{\delta (z)}{4G_2}-
i{\rm Tr}_{sc}
\left\{a(z)\epsilon^1\gamma^5d(-z)\epsilon^1\gamma^5\right\},
\nonumber\\
&& \Gamma_{\Delta\Delta^*}(z)=-\frac{\delta^2{\cal S}^{(2)}_{1}}
{\delta\Delta^{*1}(y)\delta\Delta^1(x)}=\frac{\delta (z)}{4G_2}-
i{\rm Tr}_{sc}
\left\{d(z)\epsilon^1\gamma^5a(-z)\epsilon^1\gamma^5\right\},
\label{23}
\end{eqnarray}
where $\delta (z)$ is the Dirac $\delta$-function, $z=x-y$ and
$a(z),d(z)$ are operators (\ref{121}) and (\ref{124}),
respectively. Since $\delta^2{\cal S}^{(2)}_{1}/
(\delta\Delta^1\delta\Delta^{1})=$ $\delta^2{\cal S}^{(2)}_{1}/
(\delta\Delta^{*1}\delta\Delta^{*1})=0$ and
\[
\frac{\delta}{\delta\varphi_1}=\frac 1{\sqrt{2}}
\left (\frac{\delta}{\delta\Delta}+\frac{\delta}{\delta\Delta^*}
\right ),~~
\frac{\delta}{\delta\varphi_2}=\frac i{\sqrt{2}}
\left (\frac{\delta}{\delta\Delta}-\frac{\delta}{\delta\Delta^*}
\right ),
\]
the matrix elements of the inverse propagator matrix
(\ref{23a}) have the form:
\begin{eqnarray}
 \Gamma_{11}(z)=\Gamma_{22}(z)=\frac 12\left
(\Gamma_{\Delta^*\Delta}(z)+\Gamma_{\Delta\Delta^*}(z)\right ),~~~
 \Gamma_{12}(z)=-\Gamma_{21}(z)=\frac i2\left
(\Gamma_{\Delta^*\Delta}(z)-\Gamma_{\Delta\Delta^*}(z)\right ).
\label{230c}
\end{eqnarray}
For an arbitrary function $F(z)$, it is possible to define the
Fourier-transformed one, $\overline{F}(p)$, by the relation
\begin{eqnarray}
\overline{F}(p)=\int d^4z
F(z)e^{ipz},
~~~\mbox{i.~e.}~~~~ F(z)=\int\frac{d^4p}{(2\pi)^4}
\overline{F}(p)e^{-ipz}.
\label{24}
\end{eqnarray}
Now, using (\ref{24}), it is possible to get from (\ref{23})
\begin{eqnarray}
&&\overline{\Gamma_{\Delta^*\Delta}}(p)=\frac{1}{4G_2}-
i{\rm Tr}_{sc}\int\frac{d^4q}{(2\pi)^4}
\left\{\overline{a}(q+p)\epsilon^1\gamma^5
\overline{d}(q)\epsilon^1\gamma^5
\right\},\nonumber\\
&&\overline{\Gamma_{\Delta\Delta^*}}(p)=\frac{1}{4G_2}-
i{\rm Tr}_{sc}\int\frac{d^4q}{(2\pi)^4}
\left\{\overline{d}(q+p)\epsilon^1\gamma^5
\overline{a}(q)\epsilon^1\gamma^5
\right\},
\label{25}
\end{eqnarray}
where the Fourier-transformed expressions $\bar a(q)$, $\bar
d(q)$ can be easily derived from (\ref{121}), (\ref{124}). It
follows from (\ref{25}) that
$\overline{\Gamma_{\Delta^*\Delta}}(-p)=$
$\overline{\Gamma_{\Delta\Delta^*}}(p)$. So, taking into account
(\ref{230c}), one can easily obtain the relations (\ref{23c}).

After tedious calculations we arrive at the expression
\begin{eqnarray}
&&\overline{\Gamma_{\Delta^*\Delta}}(p)=
\frac{4ip_0}{(2\pi)^4}\int\frac{d^4q}{[(p_0+q_0+\mu)^2-E^2_{p+q}]}
\left [\frac{p_0+q_0+\mu+E_q}{D_-(q_0)}+
\frac{p_0+q_0+\mu-E_q}{D_+(q_0)}\right ]-
\nonumber\\
&&-\frac{4i}{(2\pi)^4}\int\frac{d^4q}{[(p_0+q_0+\mu)^2-E^2_{p+q}]}
\left [\frac{E_q\vec p^{~\!2}+(q_0+\mu+E_q)(\vec p\cdot\vec
q)}{E_qD_-(q_0)}+
\frac{E_q\vec p^{~\!2}-(q_0+\mu-E_q)(\vec p\cdot\vec
q)}{E_qD_+(q_0)}\right ]
\label{250}.
\end{eqnarray}
Here, the notations that were introduced after (\ref{16}) are
used. Moreover, $E_q\equiv E=\sqrt{\vec q^{~\!2}+M^2}$. Note, that expression (\ref{250}) is obtained under the
assumption that $\Delta\ne 0$, i.~e. it is valid only in the 2SC
phase. In this case, the gap equation (\ref{16}) was used to
eliminate the coupling constant $G_2$ from relation (\ref{25}) in
favour of other model parameters.

Quasiparticle dispersion laws are defined by
(\ref{26b}) or, equivalently, by the two equations
$\overline{\Gamma_{ \Delta^* \Delta}} (\pm p)=0$. One can easily
see that at $\vec p^{~\!2}=0$ the first integral in RHS
of (\ref{250}) coincides with  (\ref{26}) having an
evident zero at $p_0=0$. This is just the dispersion law of a
gapless particle (NG-boson) at $\vec p^{~\!2}=0$. Let us find the
dispersion law for the NG-boson at small nonzero values of $\vec
p^{~\!2}$, i.~e. try to solve the equation $\overline{\Gamma_{
\Delta^* \Delta}} (p)=0$ at $p\to 0$. Expanding  expression
(\ref{250}) into a Taylor-series of $(p_0,\vec p)$ at the
point $(0,\vec 0)$ and taking into account only leading terms, we
have, at $\mu\ne 0$, the following equation connecting the energy and
momentum of the massless particle (in section III, it was proved that
such a particle, NG-boson, exists in the $\Delta^1$-sector of the
model):
\begin{eqnarray}
&&\overline{\Gamma_{\Delta^*\Delta}}(p)\equiv
4ip_0H(0)- \frac{4i\vec p^{~\!2}}{(2\pi)^4}\int\frac{d^4q}
{D_-(q_0)[(q_0+\mu)^2-E^2_{q}]}\left\{1-\frac{2\vec
q^{~\!2}}{3E_q(q_0+\mu-E_q)}\right\}-
\nonumber\\
&&- \frac{4i\vec p^{~\!2}}{(2\pi)^4}\int\frac{d^4q}
{D_+(q_0)[(q_0+\mu)^2-E^2_{q}]}\left\{1+\frac{2\vec
q^{~\!2}}{3E_q(q_0+\mu+E_q)}\right\}+\cdots =0,
\label{251}
\end{eqnarray}
where the quantity $H(0)$ (which is nonzero at $\mu\ne 0$) is
given in (\ref{26d}). Obviously,  the quadratic dispersion law,
$p_0\sim \vec p^{~\!2}$, for a massless particle follows from
(\ref{251}). However, in case of relativistic invariance of the
system (i.~e.{} at $\mu =0$ and for coupling constants from the region
$\omega$ (\ref{omega})), or for a color-neutral ground state, when $H(0)=0$,
the dispersion law for a NG-boson
changes. Indeed, in this case the term $4ip_0H(0)$ from
(\ref{251}) should be replaced by  $4ip^2_0H'(0)$ ($H'(0)$ is
presented in (\ref{26d})), and one arrives at a linear
dispersion law for NG-bosons.

\end{document}